\documentclass[11pt]{article}

\usepackage[utf8]{inputenc}
\usepackage[T1]{fontenc}
\usepackage{lmodern}
\usepackage{amsmath,amssymb}
\usepackage{booktabs}
\usepackage[margin=1in]{geometry}
\usepackage[round]{natbib}   
\usepackage[hidelinks]{hyperref}

\title{\bf\Large What Naturalness Measures: Fine-Tuning and Informational Invariants in Cosmology and Dark Matter}

\author{Stefano Profumo\\[2pt]
\small Department of Physics and Santa Cruz Institute for Particle Physics,\\
\small University of California, Santa Cruz, California 95064, USA\\
\small \texttt{profumo@ucsc.edu}}

\begin{document}

\maketitle

\begin{abstract}
\noindent Naturalness is commonly presented as an objective constraint on physical theories: a model requiring fine-tuning is judged implausible. This presentation conflates a representation-dependent quantity with an invariant one. A fine-tuning verdict depends on the choice of fundamental parameters, the prior, and the measure convention, so it does not by itself fix a feature of the world. Here, I argue that what is objective is structural: the universality class of the map from parameters to observables, invariant under admissible changes of parametrization and measure convention, and independent of any prior over parameter space; it constitutes an informational invariant. On this account naturalness is neither an aesthetic preference nor an objective probability, but a statement about the distinguishability geometry of the representations through which physics encodes observation. I trace the certainty of naturalness verdicts to a tradition, from Ockham through Dirac and Weinberg, in which parsimony and beauty are taken as guides to truth; modern naturalness inherits that tradition's authority without its successive justifications. The argument is developed in the gravitational and cosmological sector, where naturalness reasoning is sharpest and its effective-field-theory grounding is weakest, though that grounding remains relevant and powerful elsewhere. A uniform analysis across gravitational and particle dark matter candidates shows that fine-tuning tracks the analytic structure of the abundance map, not the nature of the candidate; that the resulting classification is invariant across measure conventions while the tuning number is not; and that this decomposition instantiates informational structural realism. I situate the position against the autonomy-of-scales account, which the argument largely accepts, and against the deflationary reading, which identifies the borrowed authority but discards the structural residue.
\end{abstract}

\section{Introduction: the two faces of a number}
\label{sec:intro}

Naturalness is among the most influential principles in recent fundamental physics, and among the least examined on quantitative and epistemological grounds. It has shaped a generation of model-building, motivated the scale and design of experiments, and licensed the dismissal of otherwise viable theories on the ground that they require fine-tuning \citep{tHooft1980, Giudice2008}. In its quantitative form it issues verdicts as numbers: a model is assigned a degree of tuning, and beyond some threshold the model is judged implausible \citep{BarbieriGiudice1988}. The principle is invoked, in short, as an objective constraint on what the world is permitted to be like.

Yet naturalness does its practical work by issuing these verdicts as numbers, and the numbers move when representational choices move. The measure on which the quantitative principle rests is, typically, a derivative of an observable with respect to input parameters; its value depends on which parameters one calls fundamental, on the prior placed over parameter space, and on the convention adopted for combining sensitivities into a single figure \citep{AndersonCastano1995, Strumia1999, Fichet2012}. A principle whose verdicts vary so radically with the analyst's coordinate choices cannot cleanly report a baseline fact about physical reality, and the confidence it lends to model-building and experimental direction is borrowed rather than earned.

This paper decomposes naturalness into a representation-dependent coordinate shell and an invariant structural core. What is preserved across admissible changes of parametrization, prior, and measure convention is the universality class of the map from parameters to observables: the analytic form, power-law or single-exponential or resonant, that determines how sharply the observable responds to its inputs. This class is an informational invariant in a precise sense; it is the content preserved across the admissible representations of a construction \citep{Profumo2026ISR}. The number is a coordinate imposed on that content.

If this is right, a further question becomes pressing: why has naturalness been received as more than the structural statement it can support? I locate the source of that surplus conviction in a long tradition, running from Ockham through Kepler and Galileo to Dirac and Weinberg, in which parsimony and beauty are taken as guides to truth \citep{Galilei1623, Wigner1960, Dirac1963, WeinbergDreams1992}. What persists across that tradition is a single disposition, that simplicity and beauty guide truth; what does not persist is any one warrant for it, the ground being rewritten by the epistemic culture of each age while the expectation it supports stands unchanged. Section~\ref{sec:genealogy} reconstructs that succession of abandoned grounds. Modern technical naturalness inherits the authority of that tradition without inheriting any of the grounds the authority was built on; the certainty that attaches to a tuning verdict is borrowed, while the structural content the verdict reports is its own.

I test the framework where the conceptual stakes are highest, moving the analysis outside the electroweak hierarchy, where naturalness is well anchored by effective field theory decoupling, into the gravitational and cosmological sector. There, a uniform analysis across gravitational relics such as primordial black holes and across particle dark matter benchmarks shows that the naturalness verdict tracks the analytic structure of the abundance map and not the physical nature of the candidate. The felt force of naturalness reasoning attaches to the coordinate shell; its legitimate objective residue is the informational invariant.

I situate this position against three others. The autonomy-of-scales account I grant in the electroweak setting and contest in the gravitational one, where the informational invariant supplies the grounding that ports \citep{Williams2015}. The candid reassessment of naturalness from within the model-building tradition I read as evidence that the principle had been carrying more epistemic weight than its grounds could bear \citep{Giudice2013, Giudice2017}. Against the deflationary critique that has done most to make naturalness an object of suspicion, I argue that it correctly identifies the borrowed authority but discards the structural residue along with it, collapsing several distinct notions of naturalness into a single aesthetic target that is easier to refute than any of them \citep{Hossenfelder2018book, Hossenfelder2018}.

The remainder of the paper proceeds as follows. Section~\ref{sec:genealogy} reconstructs the genealogy of parsimony and beauty as guides to truth. Section~\ref{sec:three-concepts} separates the three concepts the single word ``naturalness'' conceals and states the concession to the autonomy-of-scales account. Section~\ref{sec:gravity} establishes the gravitational sector as the hard case. Section~\ref{sec:measures} recasts fine-tuning measures as informational objects, and Section~\ref{sec:invariant} carries the central decomposition of naturalness into an invariant class and a representational coordinate. Section~\ref{sec:isr-positive} develops the positive account of naturalness as informational invariance, and Section~\ref{sec:williams} situates it against its rivals; Section~\ref{sec:deflation} answers the deflationary view. Section~\ref{sec:conclusion} draws the consequence.

\section{Parsimony and beauty as guides to truth: a genealogy}
\label{sec:genealogy}

The conviction that the world must be simple, and that simplicity and beauty are reliable guides to truth, has outlived every specific justification ever offered for it. The aesthetic disposition has held steady across centuries while the grounds advanced for it have been rewritten, repeatedly and incompatibly, by each successive epoch. Modern technical naturalness inherits the psychological authority of this tradition without inheriting any of the grounds on which that authority was first built.

\subsection{Ontological parsimony vs.\ mathematical harmony}
\label{sec:two-principles}

The modern naturalness intuition runs together two historically distinct ideas: ontological parsimony (the minimization of entities) and mathematical harmony (the inherent beauty of physical law). Parsimony has scholastic roots. Though attributed to William of Ockham, the canonical wording \emph{entia non sunt multiplicanda praeter necessitatem} was formulated centuries later by the Franciscan commentator John Punch in 1639 \citep{Thorburn1918}. Ockham himself restricted the principle in matters touching divine will, allowing a multiplicity of causal means if it pleased the Creator to employ them; economy was expected of creation because of the characteristics of its author, and the rule bent wherever divine sovereignty was at stake. Mathematical harmony, by contrast, has a Pythagorean and Platonic lineage: it holds that the architecture of the world is beautiful, and that beauty is prime evidence of truth. The two threads are logically independent: a universe could be ontologically extravagant yet governed by elegant relations; but modern naturalness braids them, treating a fine-tuned theory as a simultaneous offense against economy and against proportion.

\subsection{From divine economy to heuristic tool}
\label{sec:harmony-to-heuristic}

In the early modern period the grounding shifted from theological volition to mathematical architecture. For Kepler the cosmos was harmonious because constructed on geometric archetypes: his early \emph{Mysterium Cosmographicum} (1596) derived the spacing of the planetary orbits from the five Platonic solids nested within one another, and his later \emph{Harmonices Mundi} (1619) sought the harmonies of the world in musical ratios among the planetary motions \citep{KeplerWorks}; the first construction is, by modern lights, simply false, and instructive for that reason: the disposition was already capable of misleading, and already being trusted past its warrant. With Galileo the ground shifted again, toward intrinsic legibility: the book of nature was written in the language of mathematics, its characters triangles, circles, and other geometric figures \citep{Galilei1623}. By the twentieth century the metaphysical claim had become a methodological tool. Dirac promoted aesthetic value to an explicit heuristic, arguing that beauty in a theory's equations is a more reliable guide to truth than temporary disagreement with discordant data \citep{Dirac1963}. Weinberg tried to naturalize this sense, casting a physicist's taste as an instrument schooled by a ledger of past successes, and defining beauty as structural rigidity: an arrangement in which no parameter can be moved without the edifice collapsing \citep{WeinbergDreams1992}. That rigidity is the direct precursor to technical naturalness, in which a theory with ``loose parts'' or finely adjusted parameters is coded as an aesthetic failure, one with direct stakes for model-building.

\subsection{The probabilistic and sociological consolidation}
\label{sec:modern-grounds}

Two further groundings underpin the modern sensitivity metrics. The first is probabilistic, through the Bayesian model-selection apparatus: the so-called ``Occam factor'' rewards models with fewer or wider parameter spaces, since tightly adjusted parameters force a model to spread its predictive probability thinly \citep{JefferysBerger1992, MacKay2003}. I argue in what follows that this vindication remains conditioned on an ungrounded prior volume over parameter space. The second grounding is sociological: naturalness has become an internalized disciplinary norm, transmitted through training and peer review and often detached from its metaphysical pedigree \citep{Craig2023}.

The upshot is a discontinuity. Theological economy, archetypal geometry, Galilean legibility, Diracian fertility, Weinbergian rigidity, the Bayesian Occam factor, and disciplinary habit are distinct and often incompatible justifications, and the expectation that simplicity guides truth has survived the collapse of each in turn. The certainty that accompanies modern fine-tuning verdicts is a psychological inheritance whose historical estate has long been dissolved.

\subsection{The limits of the claim}
\label{sec:discontinuity}

A genealogy proves nothing about validity on its own; that a disposition has had shifting and now largely abandoned grounds does not by itself establish that it is unjustified today, and to argue so would commit the genetic fallacy. The genealogy is accordingly not offered as a refutation of naturalness. The case that modern naturalness exceeds its grounds is made elsewhere and on other terms, structurally, in what follows (Sec.~\ref{sec:gravity} and~\ref{sec:invariant}), where the gap between the confidence of a naturalness verdict and the content that can be shown to underwrite it is exhibited directly. What the genealogy supplies is that gap's explanation: the conviction is inherited, carrying the authority of a tradition with centuries of accumulated prestige while inheriting none of the grounds that prestige was successively built on.

A final qualification keeps the inheritance claim from overreaching. Modern technical naturalness did not descend from beauty-talk alone; it has its own twentieth-century roots in renormalization, symmetry, and radiative stability. This genealogy is a selective reconstruction of a recurrent epistemic disposition, not a derivation of the whole concept of naturalness from aesthetics; it explains the surplus conviction a quantitative tuning verdict carries beyond its structural content, not that structural content itself.

\section{Three concepts under one word}
\label{sec:three-concepts}

The word ``naturalness'' names at least three distinct ideas: a statement about {\it symmetry}, a statement about the {\it decoupling of scales}, and a {\it quantitative measure} that issues a number and a threshold. Failing to keep them apart is the precondition for two opposite errors: treating a fine-tuning number as though it reported a deep fact about the world, and, in reaction, dismissing the whole family of ideas as aesthetics. Both run together notions that are related but not identical, and that come apart on particular models. This section separates the three, grants the first two their due, and identifies the third, the quantitative measure, as the object of the rest of the paper. Recent surveys note the same plurality \citep{Craig2023}, and the philosophical literature has begun to distinguish a fine-tuning notion from a structural one \citep{Williams2019}; the taxonomy here refines that two-way split by setting technical naturalness apart from both.

\subsection{Technical naturalness: a statement about symmetry}
\label{sec:technical}

The most precise notion is due to 't~Hooft \citep{tHooft1980}. A small dimensionless parameter is \emph{technically natural} if setting it to zero enlarges the symmetry of the theory. The symmetry then forbids the parameter from being regenerated by quantum corrections except in proportion to itself, so its smallness is radiatively stable. The electron mass is the standard illustration: $m_e \to 0$ restores a chiral symmetry, so a small electron mass is stable under radiative corrections and requires no delicate cancellation. The strong-CP angle is a second: its smallness is made natural by promoting it to a dynamical field with a shift symmetry, the mechanism underlying the axion \citep{PecceiQuinn1977}.

Two features matter here. Technical naturalness is a sufficient condition for radiative stability, grounded in a definite structural fact: the restoration of a symmetry in a limit; and it issues no plausibility verdict on theories that lack the symmetry. It says that smallness is protected when a symmetry protects it, not that unprotected smallness is forbidden or improbable. It produces no number and sets no threshold, and of the three notions it is the least exposed to this paper's criticism, because it makes the smallest claim and grounds it in the most explicit structure.

\subsection{Autonomy of scales: a statement about decoupling}
\label{sec:autonomy}

The second notion reframes naturalness as a claim about the relation between low- and high-energy descriptions. On Williams's account, a natural theory is one whose low-energy effective description is autonomous, insensitive to the details of the unknown ultraviolet, as effective field theories generically are by virtue of decoupling \citep{Williams2015, AppelquistCarazzone1975}. The hierarchy problem is then not in the first instance a large number but a threat to that autonomy: the quadratic sensitivity of the Higgs mass to the cutoff means that low-energy physics is not, in this case, decoupled from the high-energy physics it ought to be insensitive to. Williams argues this expectation is grounded both empirically and in the structure of effective field theory, so that the absence of new physics at the Large Hadron Collider is an empirical challenge to a structural feature of quantum field theory rather than the disappointment of a preference.

This is a stronger and more defensible notion than the quantitative one with which it is sometimes conflated, and Williams himself keeps the two apart, treating the autonomy reading as the tenable residue of naturalness once the fine-tuning framing is set aside \citep{Williams2019}. I grant the account its force in the setting for which it was built. In the electroweak hierarchy the Higgs is a relevant operator, decoupling is a genuine phenomenon, and the autonomy reading has real structural content. The paper does not contest it there.

\subsection{Quantitative fine-tuning measures: a number and a threshold}
\label{sec:quantitative}

The third notion is the one that issues numbers. The Barbieri--Giudice measure and its descendants, defined in Section~\ref{sec:measures}, quantify how sensitively an observable responds to fractional changes in a theory's parameters, and a model is judged natural or tuned according to whether that sensitivity falls below or above a conventional threshold \citep{BarbieriGiudice1988, AndersonCastano1995, Strumia1999, Fichet2012, Cabrera2016}. The notion has virtues the others lack: it is operational, uniform across very different models, and permits direct comparison. It is for that reason the form in which naturalness does most of its practical work, and the form in which its verdicts are reported.

It has a corresponding liability. A large fine-tuning number is not, in itself, a statement about a symmetry or about decoupling. It is a measure of local sensitivity, and its value depends on which parameters are taken as fundamental, on the prior over parameter space, and on the convention for combining sensitivities (Section~\ref{sec:measures}). The quantitative notion is thus the least grounded of the three even as it is the most precise in appearance, and the one most readily mistaken for an objective fact. This is the notion the paper analyzes.

\subsection{How they come apart, and the concession}
\label{sec:concession}

The three are related: a technically natural parameter is one whose smallness does not require ultraviolet sensitivity, and a theory with good autonomy will tend to carry small fine-tuning numbers; but the relations are not identities, and the notions diverge on particular constructions. A parameter can be technically natural and still receive a large fine-tuning figure under a parametrization that treats a derived combination as fundamental, as the coannihilation case of Section~\ref{sec:invariant} shows; a construction can carry a small number and still sit at an improbable point under any reasonable prior, as the misalignment axion of Section~\ref{sec:axion} shows. The number, the symmetry statement, and the decoupling statement are three different things, and a careful account must say which one it means.

This matters in two directions. Against overclaiming, a large fine-tuning figure does not inherit the grounding of the symmetry or decoupling notions merely by sharing their name; the number must earn its authority on its own. Against the deflationary critique, one cannot refute naturalness wholesale by refuting its weakest reading: collapsing the three into a single aesthetic target and dispatching that target leaves the symmetry and autonomy notions untouched.

The concession follows. Of the three, the autonomy notion has the strongest claim to structural grounding, and in the electroweak hierarchy I grant it without reservation. The paper's claim is therefore not that naturalness is empty. It is that the grounding the autonomy notion supplies in the electroweak setting is absent in the gravitational and cosmological sector, where neither the symmetry nor the decoupling notion has purchase, and where naturalness is nonetheless invoked, in its quantitative form, with full confidence. That quantitative measure, applied to the relic abundance of dark matter, is the object of the rest of the paper. It is not a strawman: the quantitative measure is the carrier of the plausibility verdicts the principle delivers in practice, and it is where the gap between confidence and grounding can be made exact.

\section{The gravitational sector as the hard case}
\label{sec:gravity}

Section~\ref{sec:three-concepts} granted the autonomy account its strongest ground. In the electroweak hierarchy the Higgs mass is a relevant operator; its sensitivity to the ultraviolet is a statement about decoupling within a well-defined quantum field theory on a fixed background; and the expectation that low-energy observables be insensitive to unknown high-energy physics is, in that setting, well motivated \citep{Williams2015, AppelquistCarazzone1975}. This section argues that the grounding does not survive the move into the gravitational and cosmological sector, and that it is precisely there that naturalness reasoning is at its most confident. The mismatch between confidence and grounding is widest exactly where the principle is pressed hardest, which is what makes the sector the hard case and the right place to test the thesis.

\subsection{The cosmological constant: the most confident verdict on the least secure ground}
\label{sec:cc}

The cosmological constant is routinely presented as the most severe naturalness problem in physics. The vacuum energy receives contributions from every field at every scale; an estimate cut off at the Planck scale exceeds the observed value by some hundred and twenty orders of magnitude, and even a cutoff at the electroweak scale leaves a discrepancy of sixty \citep{WeinbergCC1989, Martin2012}. As a dimension-zero, maximally relevant operator, the vacuum energy is more ultraviolet-sensitive than the Higgs mass, and on a first reading the autonomy account should therefore apply \emph{a fortiori}.

The first reading is mistaken, and the reason is the heart of this section. The autonomy argument is internal to quantum field theory on a fixed, non-dynamical spacetime; decoupling is a theorem about how heavy fields fail to influence low-energy observables within such a theory \citep{AppelquistCarazzone1975}. The cosmological constant escapes that framework in two connected respects. First, in quantum field theory without gravity the absolute value of the vacuum energy is unobservable; only energy differences enter, and the vacuum energy acquires physical significance only when it gravitates, that is, only when it sources curvature through the Einstein equations. The problem therefore lives at the interface between field theory and gravity, not inside the decoupling structure of a field theory. Second, to judge whether the observed value is natural one must know how vacuum energy couples to gravity across all scales, including the Planckian regime where the effective description of gravity itself breaks down. The Wilsonian picture that underwrites Higgs naturalness presupposes a controlled procedure for integrating out heavy modes within a well-defined theory; for the cosmological constant the modes to be integrated out include gravitational physics for which no controlled effective theory is available. The decoupling statement that grounds naturalness for the Higgs simply does not exist for the cosmological constant.

This is not to reduce the problem to a complaint about the size of a number, and the distinction is worth stating once, clearly. The failure of decoupling is a structural pathology, not merely the absence of a structural comfort: quantum field theory dictates ultraviolet-sensitive contributions at every scale, and once the constant gravitates there is no mechanism within the framework that stabilizes the observed value short of a cancellation to one part in $10^{120}$. That is a formal breakdown of predictive control at the field-theory/gravity interface, and it should be distinguished from a preference for tidy numbers. But the breakdown is a breakdown of the very EFT reasoning that grounds the Higgs verdict; it removes the conditions under which an autonomy grounding could be stated, rather than supplying one. And the quantitative claim (that a value of order $10^{-120}$ is thereby improbable) still asserts more than the diagnosis licenses: the diagnosis says the framework loses control, not that the controlled answer would have been of order unity. This is what separates the cosmological constant from the relic-abundance maps that occupy the rest of the paper, where the physics is under control and the tuning is a clean property of cosmological dynamics rather than a symptom of the framework breaking down.

The other two lenses sharpen rather than soften the verdict. For example, the Barbieri--Giudice measure (Sec.~\ref{sec:measures}), applied here, returns an enormous sensitivity: the residue is a difference of contributions each of order the fourth power of its scale, so the logarithmic derivative of the residue with respect to any one contribution is of order $10^{120}$. But this is evaluated at an additive cancellation, and a multiplicative log-derivative is inflated without bound near an additive zero, for the same chart-internal reason that deflates the coannihilation splitting (see Sec.~\ref{sec:tier-invariant}). The colossal figure is a representation-dependent coordinate of the most extreme kind, {\it not an invariant verdict}. The probabilistic reading associated with Weinberg makes the missing ingredient explicit: to call the value improbably small is to weigh it against a distribution over the values the vacuum energy might have taken, and the anthropic bound supplies exactly that distribution: a measure over a landscape, cut by the requirement that structure form, against which the value is rendered typical \citep{Weinberg1987}. The move concedes that the problem is at bottom probabilistic, and lodges the entire verdict in a prior over universes that an embedded observer cannot ground (as I argue in detail in Sec.~\ref{sec:embedded}).

The situation is thus the reverse of the rhetoric. The cosmological constant is at once the most dramatic naturalness problem, by the size of the number, and the case in which the autonomy grounding is least secure, because securing it would require extending decoupling reasoning across a scale at which that reasoning fails. None of the three frameworks delivers a grounded verdict that the cosmological constant is unnatural: decoupling cannot be stated across the Planck scale, the Barbieri--Giudice number is a coordinate inflated by an additive cancellation, and the probabilistic reading only relocates the question into a prior an embedded observer has no way to fix. I do not claim that no structural approach exists: for instance, unbroken supersymmetry would forbid a vacuum energy, and sequestering and unimodular formulations remove the constant from the gravitating stress-energy by construction. The point is narrower and, I think, decisive: none of these supplies the clean decoupling statement the autonomy account requires, and so none converts the verdict into the kind of grounded structural claim the same account delivers for the Higgs.

\subsection{The anthropic turn as a symptom}
\label{sec:anthropic}

The history of the two problems is itself evidence for the asymmetry. When naturalness came under pressure in the electroweak sector, the response was structural model-building: supersymmetry, technicolor, and composite-Higgs constructions were advanced precisely because the autonomy grounding pointed toward concrete ultraviolet completions at an accessible scale \citep[see e.g. the discussion in][]{Giudice2008}. The expectation made a prediction, new physics near the TeV scale, and the failure of that prediction at the Large Hadron Collider was felt as an empirical result because the expectation had structural content to begin with.

The cosmological constant admitted no such response. There was no autonomy grounding to suggest a structural completion, and the community turned instead to anthropic selection within a landscape of vacua \citep{Weinberg1987, BoussoPolchinski2000, Susskind2003}. The anthropic turn is not a structural explanation of the observed value; it is what remains when the naturalness expectation fails and no decoupling-based rescue is available. Where the grounding was present, its failure provoked further structural physics; where the grounding was absent, its failure provoked a change in what counts as explanation. The structural constructions the cosmological constant did prompt never amounted to a clean decoupling-based completion at an accessible scale, the way supersymmetry and compositeness did for the Higgs; there was never an autonomy expectation there to be vindicated, only an aesthetic one to be disappointed.

\subsection{Dark energy: the invariant relation and its revisable encoding}
\label{sec:dark-energy}

The cosmological constant is a live problem, not only a historical one, and its current status sharpens the point in the language of the companion account \citep{Profumo2026ISR}. The observed late-time acceleration is encoded in the standard cosmological model as a constant vacuum energy. Recent data from the Dark Energy Spectroscopic Instrument prefer instead an evolving equation of state, with a departure from the constant value at a significance ranging from roughly three to four standard deviations depending on which supernova compilation is included \citep{DESI2025}. What the data secure is not the cosmological constant as such, but the shape of the distance--redshift relation as jointly constrained by baryon acoustic oscillations, the microwave background, and supernovae. That relation is the informational invariant; whether it is best encoded by a cosmological constant, a two-parameter evolving equation of state, a dynamical scalar field, or something else is a representational question under active revision, and the significance of the departure shifts with the choice of supernova catalog, with the inference pipeline rather than with the sky \citep{Profumo2026ISR}.

The decomposition can be made precise in the same terms as the abundance map, and the precision is worth the detour, because it shows the split operating on the flagship problem with the apparatus the rest of the paper uses. The distance--redshift relation is an integral functional of the expansion history, $D(z) = \int_0^z dz'/H(z')$, and the dark energy enters $H$ only through $\rho_{\rm DE}(a) = \rho_{\rm DE,0}\,\exp\!\big[\,3\!\int_a^1 (1+w(a'))\, d\ln a'\,\big]$. The observable is therefore a double integral of the equation of state, a smoothing that renders large families of histories $w(a)$ nearly indistinguishable in the data. Parametrize $w(a) = w_0 + w_a(1-a)$, the form the recent analyses adopt \citep{DESI2025}; the Fisher information of the distance data on $(w_0, w_a)$ is, exactly as for the abundance map of Section~\ref{sec:measures}, a geometry with a stiff direction and a sloppy one. The stiff direction is the value of the equation of state at the pivot redshift where the data constrain it best, $w_p \equiv w(a_p)$; the sloppy direction is the rate of evolution, the combination the smoothing integral suppresses. The invariant content is $w_p$, equivalently the distance--redshift relation it summarizes; the split into a present value $w_0$ and a derivative $w_a$ is a chart on that stiff--sloppy geometry, decorrelated only at the pivot.

This is why the reported significance of the departure from a cosmological constant moves with the supernova compilation. The constant, $w \equiv -1$, sits in the $(w_0, w_a)$ chart at $(-1, 0)$, and the displacement of the data's preferred point from $(-1, 0)$ is read largely along the sloppy direction, where the likelihood is shallow and the prior and choice of catalog do much of the work; that is the direction along which a three-sigma and a four-sigma verdict differ. The stiff direction, $w_p$, the part every adequate analysis agrees on, barely moves between compilations. The naturalness of the cosmological constant is thus a claim about the location of one encoding along the sloppy, representation-dependent direction, while the secure content, the distance--redshift relation captured by $w_p$, is the stiff invariant. The invariant-versus-representational decomposition I present in what follows thus appears already at the level of the flagship problem, in the same distinguishability geometry, before any abundance map is written down.

\subsection{From vacuum energy to relic abundance}
\label{sec:to-abundance}

The third arena of gravitational-sector naturalness is the one this paper analyzes in detail: the relic abundance of dark matter. It differs from the cosmological constant in that the relevant observable, the present-day abundance $\Omega_{\rm DM} h^2$, is a sharply measured number that every candidate must reproduce \citep{Planck2018}; it differs from dark energy in that the encoding question is posed not about a single background quantity but about an entire landscape of possibler dark matter production mechanisms. What it shares with both is the feature that makes the sector the hard case. The abundance is fixed by cosmological dynamics: by thermal and non-thermal processes in the early universe and, for gravitational relics such as primordial black holes, by gravitational collapse \citep{Profumo2026PBH}. None of these is a decoupling phenomenon, and the autonomy account that grounds naturalness for the Higgs has no straightforward purchase on any of them. Yet verdicts are issued here with full confidence: that primordial black holes are fine-tuned, that the thermal relic with weak-scale mass and coupling is natural. The remainder of the paper takes the relic abundance as its worked observable and asks what, in those confident verdicts, is invariant and what is borrowed. The autonomy account is engaged directly in Section~\ref{sec:williams}; the intervening sections establish the measure and extract the invariant.

\section{What the measures measure}
\label{sec:measures}

Quantitative naturalness begins with a measure of sensitivity. Barbieri and Giudice, seeking to express how strongly the electroweak scale depends on the parameters of a supersymmetric spectrum, defined
\begin{equation}
\Delta \;\equiv\; \max_i \Delta_i , \qquad
\Delta_i \;\equiv\; \left| \frac{\partial \ln \mathcal{O}}{\partial \ln x_i} \right| ,
\label{eq:bg}
\end{equation}
where $\mathcal{O}$ is a physical observable a construction must reproduce and $\{x_i\}$ are the independent parameters the theory takes as inputs \citep{BarbieriGiudice1988}. In the original application $\mathcal{O}$ is the electroweak scale and the $x_i$ are soft supersymmetry-breaking masses and couplings; in the cosmological application here $\mathcal{O}$ is the dark matter relic abundance $\Omega_{\rm DM} h^2$ and the $x_i$ are the masses, couplings, and cosmological initial data a given production mechanism specifies independently. The derivative is evaluated on the contour where the prediction meets the observed value, so $\Delta_i$ records how stiffly the input must be held once the observable is fixed: $\Delta_i = n$ means a fractional change $\epsilon$ in $x_i$ produces a fractional change $n\epsilon$ in $\mathcal{O}$. A construction is deemed natural when $\Delta$ is of order unity and fine-tuned when $\Delta$ is large. Several features of this definition, usually treated as technical caveats, are in fact the philosophically decisive ones.

First, the measure is a logarithmic derivative, and the Strumia--Rattazzi variant makes its geometric character explicit \citep{Strumia1999}:
\begin{equation}
\Delta_{\rm SR} \;=\; \Big( \sum_i \Delta_i^2 \Big)^{1/2}
\;=\; \big| \nabla_{\ln x}\, \ln \mathcal{O} \big| ,
\label{eq:sr}
\end{equation}
the magnitude of the gradient of the log-observable in log-parameter space. That this is a Fisher-information length, and not merely an evocative redescription of a gradient, follows once one notes that the abundance is not a bare deterministic map but a measured quantity. Planck fixes $\Omega_{\rm DM} h^2 = 0.1200 \pm 0.0012$, a Gaussian likelihood whose fractional width is $r \equiv \sigma_\Omega / \Omega \approx 10^{-2}$ \citep{Planck2018}. Pulling that likelihood back through the map $x \mapsto \mathcal{O}$ induces a Fisher metric on parameter space,
\begin{equation}
g_{ij} \;=\; \frac{1}{\sigma_\Omega^2}\,
\frac{\partial \mathcal{O}}{\partial x_i}\,
\frac{\partial \mathcal{O}}{\partial x_j},
\label{eq:fisher}
\end{equation}
the Fisher information of the parameters under the relic-abundance measurement. The identification is exact under three idealizations: a single observable, a Gaussian likelihood, and a deterministic parameter-to-observable map. Because there is a single observable the metric has rank one; its one non-vanishing eigenvalue, in the logarithmic coordinates of Eq.~\eqref{eq:sr}, is $\Delta_{\rm SR}^2 / r^2$, and the information the measurement carries about an individual parameter is $\mathcal{I}(\ln x_i) = (\Delta_i / r)^2$. The Barbieri--Giudice sensitivity is therefore, exactly, the Fisher-information length of the parameter under the abundance measurement, in units of that measurement's fractional precision: a determination of $\Omega$ to fractional precision $r$ resolves $\ln x_i$ to a width $\sigma_{\ln x_i} = r / \Delta_i$. The number is an inferential resolution in the strict information-theoretic sense, not by analogy; and the relevant ensemble, the sampling distribution of the measurement, is one an embedded observer possesses, unlike the prior over parameters (see Sec.~\ref{sec:embedded}). The companion account adds the generalization, not the foundation: its Fisher--Rao apparatus lives on the full space of probability distributions over the cosmological data and treats many correlated observables at once \citep{Profumo2026ISR}, and the abundance measure of Eq.~\eqref{eq:fisher} is its rank-one restriction to the single projection $\Omega_{\rm DM} h^2$. The restriction is what makes the analysis tractable and uniform across candidates; the embedding in the larger geometry is what ties it to the cross-representational invariants.

Second, the measure is not invariant under reparametrization. Whether one differentiates with respect to a fundamental Lagrangian parameter or to a derived combination changes the value of $\Delta$, sometimes by more than an order of magnitude. The verdict therefore presupposes a choice of which parameters are fundamental, a choice not fixed by the low-energy theory but imported from an assumed ultraviolet completion. That the tuned quantity inherits this dependence is not peculiar to cosmology; the renormalized-parameter analysis of the Higgs reaches the same conclusion, that the need to tune the renormalized Higgs mass is to a large extent an artifact of scheme and parameter choice rather than a scheme-independent fact \citep{RosalerHarlander2019}. I adopt throughout the convention that the fundamental parameters are the independently specified inputs to the Lagrangian or to the cosmological initial conditions, evaluated at the relevant scale \citep{Profumo2026PBH}. The convention is defensible, but it is a convention, and the dependence of the verdict upon it is the first sense in which the number is representational.

Third, the measure is local. It reports the sensitivity of the observable at a point, not the volume of viable parameter space, nor the prior probability of arriving at the observed value. These can diverge sharply: a construction can be locally insensitive, with small $\Delta$, yet occupy a region improbable under any reasonable prior. Bayesian naturalness measures address the volume question directly, at the cost of an explicit prior \citep{AndersonCastano1995, Fichet2012, Cabrera2016}; the relation between the local measure and the global one is itself a place where representational choices enter, and the misalignment axion of Section~\ref{sec:axion} will show how far the two can come apart.

The observable completes the setup. For the remainder I take $\mathcal{O}$ to be $\Omega_{\rm DM} h^2$, fixed by observation to $0.120$ \citep{Planck2018}. The choice is deliberate: the relic abundance is a single, sharply measured number every candidate must reproduce, whatever its nature, and therefore furnishes a common target against which gravitational relics and particle candidates can be compared under one measure. It is also the point at which the analysis becomes specific to the gravitational sector, since the abundance is set by cosmological dynamics, and for gravitational relics by gravitational collapse, a regime in which the autonomy grounding of Section~\ref{sec:three-concepts} has no straightforward purchase. A measure of sensitivity, a convention fixing its parameters, a locality that separates it from probability, and a single cosmological observable: these are the materials. The following section~\ref{sec:invariant} shows what, in what they produce, is invariant.

\section{The invariant and the representational}
\label{sec:invariant}

Above we recast the fine-tuning measure as a geometric object: the Barbieri--Giudice and Strumia--Rattazzi quantities are distinguishability lengths on the map from inputs $\{x_i\}$ to the observable $\mathcal{O} = \Omega_{\rm DM} h^2$. Granting that, this section asks the question a philosophical reading of naturalness must answer: of the fine-tuning a construction is assigned, which part is a feature of the world and which a feature of the description we impose on it?

One clarification fixes the scope of the claim at the outset. The invariance asserted below is invariance under change of \emph{representation of a fixed map} (reparametrization, prior, and measure convention), not invariance under change of the \emph{physical model} that determines the map. Better physics can redraw the abundance map and hence the class. The two kinds of change are different, and the present account locates objective content in the structure that survives the first.

\subsection{The candidate does not set the tuning; the map does}
\label{sec:map-not-candidate}

The distinction the classification rests on is one of functional form. The measure of Eq.~\eqref{eq:bg} reads the logarithmic response of the observable to its controlling input, $\Delta = \lvert \partial \ln \mathcal{O}/\partial \ln x \rvert$, so what governs $\Delta$ is the order of growth of the abundance map $\mathcal{O}(x)$ near the contour on which the observed value is reproduced. By that order of growth the maps fall into three universality classes:
\begin{itemize}
\item[] \emph{Class~I}: maps with no exponential factor, whose leading dependence on the controlling input is algebraic, $\mathcal{O} \sim x^{n}$; the sensitivity $\Delta = \lvert n \rvert$ is the power-law exponent, set by scaling dimensions and independent of where on the map the observed value falls.
\item[] \emph{Class~II}: maps dominated by a single exponential, $\mathcal{O} \sim e^{-f(x)}$ with $f$ log-linear in the controlling input to leading order; the sensitivity is whatever value of $f$ the observed-abundance contour demands, $\Delta = \lvert \partial f/\partial \ln x \rvert = O(f)$, fixed by a ratio of physical scales rather than by the candidate.
\item[] \emph{Class~III and beyond}: maps whose leading growth is steeper than a single exponential (such as resonant, cancellation-driven, or double-exponential) for which $f$ grows faster than log-linearly and $\Delta$ carries no contour-fixed ceiling.
\end{itemize}
The classification is by functional form alone. The numerical ranges that accompany each class in Table~\ref{tab:classes} ($\Delta \lesssim 5$ for Class~I, $\Delta \approx 15\text{--}50$ for Class~II, $\Delta \gtrsim 10^{3}$ for Class~III) are consequences of these forms once the relevant scale ratios are inserted, not independent criteria; because the controlling exponents are fixed by scale ratios rather than by a continuum of adjustable inputs, the classes occupy separated bands rather than a smear. Two elementary maps make the lower classes concrete: $\mathcal{O}(x) = x^{n}$ gives $\Delta = n$, a constant fixed by the exponent; $\mathcal{O}(x) = e^{-x}$ gives $\Delta = x$, so on the observed-abundance contour the sensitivity is whatever exponent that contour demands, generically a few tens. The first map carries its tuning number in its shape, the second inherits it from the location of the contour.

When a single measure and a single observable target, $\Omega_{\rm DM} h^2 = 0.120$, are applied uniformly across a broad landscape of dark matter candidates, the fine-tuning of each construction is fixed by the analytic form of its abundance map and not by the nature of the candidate \citep{Profumo2026PBH}. Across constructions spanning gravitational relics (primordial black holes formed by biased domain walls, by an early matter-dominated era, by a first-order phase transition, and by inflationary collapse) and particle candidates (thermal-relic WIMPs on and off resonance and in coannihilation, freeze-in, asymmetric dark matter, and axions in both misalignment and post-inflationary histories), all three classes are realized; Table~\ref{tab:classes} collects representative cases.

\begin{table}[t]
\centering
\small
\begin{tabular}{@{}llcc@{}}
\toprule
Construction & Production type & Map class & $\Delta$ \\
\midrule
Asymmetric dark matter        & particle             & I (power law)        & $1$ \\
Post-inflationary axion       & particle             & I (power law)        & $1.19$ \\
Misalignment axion            & particle             & I (power law)        & $\approx 2$ \\
Off-resonance thermal WIMP    & particle             & I (power law)        & $\lesssim 5$ \\
Freeze-in                     & particle             & I (power law)        & $\lesssim 5$ \\
Biased-domain-wall PBH        & gravitational relic  & I (power law)        & $2\text{--}4.5$ \\
\midrule
Coannihilating WIMP           & particle             & II (single exp.)     & $\approx 48$ \\
Early-matter-domination PBH   & gravitational relic  & II (single exp.)     & $15\text{--}50$ \\
Phase-transition PBH$^{\dagger}$ & gravitational relic & II (single exp.)   & $15\text{--}50$ \\
\midrule
Higgs-funnel (resonant) WIMP  & particle             & III (resonant)       & $\approx 6500$ \\
Ultra-slow-roll inflation PBH & gravitational relic  & III (double exp.)    & $\gtrsim 10^3$ \\
\bottomrule
\end{tabular}
\caption{Representative dark matter constructions, their production type, the
universality class of the abundance map, and a representative Barbieri--Giudice
sensitivity $\Delta$ evaluated in an admissible parametrization on the
observed-abundance contour \citep{Profumo2026PBH}. The class is fixed by the
analytic form of the map and is orthogonal to whether the candidate is a particle
or a gravitational relic: every production paradigm contributes to more than one
class, and the primordial-black-hole paradigm alone spans all three. Within a class
the number varies with benchmark and parametrization: the coannihilation entry
moves by a factor of forty under the inadmissible $\delta$ chart
(Section~\ref{sec:tier-invariant}), while the class does not.\\
$^{\dagger}$Conditional on the single-exponential approximation; a more faithful
super-exponential collapse probability raises this construction to Class~III
(see Sec.~\ref{sec:single-exp}). The table displays the representative landscape that
motivates the philosophical claim; model-by-model derivations are in
\citet{Profumo2026PBH}.}
\label{tab:classes}
\end{table}

The single-exponential band sits at the large exponent $\ln(\Omega_{\rm natural}/\Omega_{\rm DM}) \approx \ln(T_{\rm form}/T_{\rm eq})$ (see Sec.~\ref{sec:single-exp}), of order ten to fifty, while the power-law band sits at order unity, and the formation scales realized across the landscape leave the intervening range thinly populated. The philosophical lesson follows from this band structure rather than from any one number. Had $\Delta$ filled its range smoothly, the tier would be an artifact of where one drew the bin edges; that it clusters by analytic form is what marks the class, and not the number, as the invariant. The two most natural constructions are particle candidates, asymmetric dark matter ($\Delta = 1$) and the post-inflationary axion ($\Delta = 1.19$); the most tuned construction whose measure can be computed precisely is also a particle, the Higgs-funnel WIMP ($\Delta \approx 6500$). Biased-domain-wall primordial black holes sit at $\Delta = 2\text{--}4.5$, in Class I, alongside the off-resonance WIMP and freeze-in.

The reading is immediate. The familiar verdict that primordial black holes are a fine-tuned dark matter candidate is a category error. The paradigm spans all three classes, from the gravity-fixed domain-wall construction at $\Delta = 2$ to single-field ultra-slow-roll inflation at $\Delta \gtrsim 10^3$; the worst case has been mistaken for the paradigm. Fine-tuning is not a property of whether dark matter is a particle or a gravitational relic. It is a property of the analytic structure of the map by which model parameters determine the abundance.

\subsection{The tier is invariant; the number is a coordinate}
\label{sec:tier-invariant}

The Barbieri--Giudice measure is not invariant under reparametrization, and this is the crux of the matter. Consider WIMP coannihilation. Parametrized by the fractional mass splitting $\delta = (M_{\rm NLSP} - M_\chi)/M_\chi$, the measure evaluates to $\Delta_\delta \approx x_F \delta \approx 1.3$, reading as nearly perfect naturalness. Parametrized instead by the two independent soft masses $(M_\chi, M_{\rm NLSP})$, the same physics yields $\Delta_{M_{\rm NLSP}} \approx 48$. One construction, one observable, two parametrizations, a factor of forty between the verdicts: the number is not a fact about the world but a fact about the coordinates imposed on it.

If a change of coordinates can move the number, one might conclude that nothing it tracks is invariant. The invariance claim must therefore be made against a defined class of coordinate changes. The decisive criterion is a regularity condition on the chart the measure actually reads, and it can be checked on the map alone, before any tuning number is computed. The measure of Eq.~\eqref{eq:bg} is a \emph{logarithmic} derivative. Call a reparametrization \emph{admissible} when it is a smooth, invertible change of inputs whose Jacobian, taken in the logarithmic coordinates the measure reads, is finite and non-vanishing on a neighborhood of the observed-abundance locus. This is the standard regularity condition any change of coordinates must meet to be a local diffeomorphism, and it makes no reference to the tier or the tuning number.

The $\delta$ chart fails it, and the failure is purely chart-internal. The linear change $(M_\chi, M_{\rm NLSP}) \leftrightarrow (M_\chi, \delta)$ is itself perfectly regular, with Jacobian $1/M_\chi$ finite and non-zero even at $\delta = 0$; the pathology is not a defect in the differentiability of $\delta$. It is that $\ln\delta$ is singular at the evaluation point: $\partial \ln \delta / \partial \ln M_{\rm NLSP} = (1+\delta)/\delta$ diverges as $\delta \to 0$, and $\delta$ vanishes precisely at the mass-degeneracy locus where the observed abundance is reproduced. A multiplicative log-derivative evaluated where its own coordinate has a zero is deflated for a reason internal to the chart and external to the physics of the map. The soft masses $(M_\chi, M_{\rm NLSP})$, specified at the mediation scale, carry no such singularity in their logarithmic chart; they are admissible, and they return the construction to its tier. The cosmological constant's $10^{120}$ figure (Sec.~\ref{sec:cc}) is the same pathology in its most extreme form: a log-derivative inflated, rather than deflated, at an additive cancellation.

A secondary remark answers a residual worry. One might add to admissibility the requirement that the coordinates be quantities the theory specifies \emph{independently} as inputs at a fixed defining scale such as the Lagrangian couplings and soft masses, or the cosmological initial data, which $\delta$, a derived combination of the two soft masses, is not. This input-status condition is a useful diagnostic, and it agrees with the regularity criterion on every case in Table~\ref{tab:classes}; but it is the weaker of the two, because which quantities a theory ``specifies independently'' can itself shift between ultraviolet completions. The regularity criterion does not: it is settled on the chart and the map, with no appeal to a preferred completion. I therefore rest the invariance on regularity and treat input-status as corroborating. Admissibility, so understood, does not banish $\delta$ from physics; the coannihilation threshold is a real mechanism, and $\delta$ is the natural variable in which to express it. It bars $\delta$ only from the single role of independent input to the sensitivity measure, the role in which a constrained combination would be treated as a freely dialed parameter.

The decomposition is then exact, and it is a theorem rather than a stipulation. The tier is the order of growth of $\mathcal{O}(x)$ near the observed-abundance locus, a property of the function's germ. An admissible reparametrization is a local diffeomorphism with non-singular logarithmic Jacobian, and the order of growth of a function is preserved under such a map: a power law remains a power law and an exponential an exponential, because a finite, non-vanishing Jacobian rescales the gradient without altering the functional class. The number $\Delta$ is the magnitude of that gradient in a chosen chart and transforms by the Jacobian; the tier is the diffeomorphism-invariant content, and the number is a coordinate on it. This licenses the central definition:
\begin{quote}
\emph{A naturalness universality class is an equivalence class of local abundance
maps $\mathcal{O}: x \mapsto \Omega$, taken under admissible reparametrizations of
the input coordinates (smooth, invertible changes whose logarithmic Jacobian is
finite and non-vanishing near the observed-abundance contour), that preserve the
local order of growth of $\ln \mathcal{O}$ on that contour: algebraic (Class~I),
single-exponential (Class~II), or resonant, super-exponential, and higher (Class~III).}
\end{quote}
Benchmark, parametrization, and measure convention vary freely within a class, and the tuning number is the coordinate they move.

A second invariance runs alongside the first. Just as the tier is stable under admissible changes of parametrization, it is stable under changes of measure convention: across the Barbieri--Giudice measure, the Strumia--Rattazzi quadrature measure, and the island half-width, no construction in the landscape changes class \citep{Profumo2026PBH}. Two independent sources of representational freedom, the choice of fundamental parameters and the choice of measure, both leave the class fixed while moving the number. This is the decomposition that informational structural realism prescribes \citep{Profumo2026ISR}: objective content is located not in the quantities that vary from one representation to another, but in the structure preserved across them. The objective content of a naturalness claim is the class, not the number.

\subsection{A cosmological-clock invariant within the tuned regime}
\label{sec:single-exp}

The single-exponential class furnishes the sharpest illustration, because it contains an invariant even where the language of naturalness would lead one to expect only contingent ugliness. For any construction whose abundance takes the single-exponential form, the measure evaluated on the observed-abundance contour satisfies
\begin{equation}
\Delta \;\Big|_{\Omega = \Omega_{\rm DM}}
= \ln\!\left( \frac{\Omega_{\rm natural}}{\Omega_{\rm DM}} \right)
\;\approx\; \ln\!\left( \frac{T_{\rm form}}{T_{\rm eq}} \right)
\;\approx\; 14\text{--}50 ,
\label{eq:single-exp-identity}
\end{equation}
where $\Omega_{\rm natural}$ is the order-unity density the construction would yield with the exponential suppression switched off, $T_{\rm form}$ is the temperature at which the relic forms or decouples, and $T_{\rm eq} \simeq 0.8$~eV is the temperature at matter--radiation equality; the identity is independent, up to an order-unity elasticity with which the microscopic input controls the exponent, of the microphysics that supplies the exponential \citep{Profumo2026PBH}; the derivation, and the precise role of that elasticity, are in Appendix~\ref{app:single-exp}. For a first-order phase transition the result is independent of the function $S(\alpha)$ that encodes the transition dynamics; for early matter domination and for coannihilation the same identity holds with the appropriate formation or freeze-out scale. The tuning of Class II is therefore not an accident of any particular model. It is forced by the ratio of two physical scales: the scale of formation or freeze-out and the scale of matter--radiation equality; it is set by a cosmological clock, by gravity and cosmology, not by the identity of the candidate. A phase-transition primordial black hole, an early-matter-domination primordial black hole, and a coannihilating WIMP share a tier for this structural reason, despite spanning modulus decay, hidden-sector percolation, and electroweak thermodynamics.

An honest caveat strengthens the point. The Class II placement of the phase-transition primordial black hole is conditional on the single-exponential approximation; a more accurate treatment of the collapse probability is super-exponential and raises the construction to Class III \citep{Profumo2026PBH}. The tier can migrate when the analytic structure of the map is described more faithfully, and this is exactly what the account of Section~\ref{sec:invariant} predicts: the invariant content lives in the structure of the map, so refining the map can move the class. What cannot be rehabilitated is the idea that the number was ever the thing being tracked.

\subsection{The axion: prior probability versus local sensitivity}
\label{sec:axion}

The misalignment axion exhibits the naturalness distinction and the informational distinction at once, and is the cleanest fusion of the two halves of this programme. Its relic abundance is a power law,
\begin{equation}
\Omega_a h^2 \;\simeq\; 0.18\, \theta_i^2
\left( \frac{f_a}{10^{12}\ \mathrm{GeV}} \right)^{1.19} F(\theta_i),
\label{eq:axion}
\end{equation}
so the local measure is $\Delta \approx 2$ for a misalignment angle $\theta_i \sim 1$. The same local value obtains at $\theta_i \approx 3 \times 10^{-3}$, the value required when $f_a$ is pushed to a grand-unified or string scale, because the power-law structure of Eq.~\eqref{eq:axion} is unchanged by the benchmark. Local sensitivity is invariant under the choice of benchmark.

Yet $\theta_i \approx 3 \times 10^{-3}$ is a factor of roughly $300$ below the natural value of order unity, and on a flat prior over $[-\pi, \pi]$ the probability of landing there is of order $10^{-3}$. This prior tuning of $\sim 300$ is a real cost, qualitatively different from, and invisible to, the local measure. The two quantities map onto the invariant and the representational. The local sensitivity is the stiff, likelihood-dominated, prior-robust direction that any admissible representation must agree on; the prior tuning is the sloppy, prior-driven, representation-dependent quantity that depends on a modeling choice, here a flat prior on the misalignment angle. One object thereby displays both distinctions simultaneously: the naturalness distinction between local sensitivity and prior probability, and the informational distinction between what is invariant and what is representational.

The example also explains a fact the language of naturalness renders puzzling. A generic axion with an order-one misalignment angle and an anthropically selected axion with a tiny one receive the same verdict from the invariant measure, because that measure is, correctly, blind to the prior, and the prior is where the difference between the two situations lives. The verdict is not defective; it is reporting the invariant and declining to report the coordinate.

\medskip\noindent
That separation is now in hand, and it organizes the rest of the paper. Naturalness has an invariant core, the universality class of the abundance map, or, equivalently, the distinguishability structure any admissible representation must preserve, and a representational shell: the tuning number, the choice of fundamental parameters, the measure convention, and the prior. The discontinuity this paper presses can now be stated precisely. The authority the older tradition of parsimony and beauty confers attaches to the shell, presenting a representation-dependent coordinate as though it were a deep fact about how contrived the world is. The defensible content is the invariant class, which is informational and structural, not aesthetic. 

\section{Naturalness as informational invariance}
\label{sec:isr-positive}

This section turns the decomposition into a positive account: it says what naturalness is, once the borrowed authority is set aside, by identifying the invariant core with an informational invariant of the kind the companion account locates at the base of cosmological knowledge \citep{Profumo2026ISR}.

\subsection{The positive thesis}
\label{sec:positive-thesis}

The structural realist tradition holds that what survives theory change most securely is not the inventory of objects a theory posits but the structure it encodes \citep{Worrall1989, Ladyman1998}. The companion account sharpens this for cosmology by locating objective content not in the structure internal to any one successful representation but in the structure preserved across empirically adequate representations: cross-representational informational invariants, characterized through correlation structure, distinguishability geometry, and bounds on accessible information \citep{Profumo2026ISR}. The objective content of naturalness is an invariant of exactly this kind: the universality class of an abundance map, an equivalence class under admissible reparametrization that shares the same analytic structure and therefore the same distinguishability geometry, regardless of which parameters are called fundamental or which convention assigns the number. Section~\ref{sec:tier-invariant} showed that this class is stable across measure conventions while the number is not; that stability is what it means for the class to be a cross-representational invariant. Naturalness, on this account, is a claim about the distinguishability geometry of the map from parameters to a cosmological observable, and its objective content is the equivalence class that geometry defines. It is neither a statement about the elegance of the world nor an objective probability over it.

This is where the account must answer the standing objection to structural realism. Newman's challenge is that structure, taken abstractly, is cheap: any domain of suitable cardinality instantiates any structure, so a doctrine that only structure is known threatens to reduce to a claim about cardinality together with a free choice of relations \citep{Ladyman1998}. The objection has force against a structuralism whose structure is a bare set-theoretic relation, ascribed at will. It has none against the structure at issue here, because that structure is not ascribed but measured. The distinguishability geometry of Section~\ref{sec:measures} is the Fisher geometry pulled back from an actual likelihood, individuated by which observable resolves which parameter direction and to what precision; a universality class is not a relation any model trivially satisfies but the empirically constrained answer to how the measured abundance responds to its parameters. Newman's collapse is blocked by the very fact that makes the invariant informational rather than aesthetic: it is anchored to a measurement, not to cardinality. It is worth being exact about what this blocks. Newman's objection is aimed at ontology, at the thesis that reality possesses nothing but structure, whereas the measured-likelihood reply speaks in the first instance about a representation. What it establishes is that an embedded observer's access to the parameter has a definite, non-arbitrary structure, individuated by which observable resolves which direction and to what precision, and so is not the freely ascribed cardinality-plus-relations the collapse trades on. That is a claim about epistemic access, and it is the claim the informational reading needs. It does not by itself establish that the parameter is nothing over and above that structure (the ontic thesis) and it should not be read as quietly supplying it; whether the account licenses the ontic upgrade is taken up in Section~\ref{sec:realism-status} below.

\subsection{Inferential resolution, not ontological contrivance}
\label{sec:resolution}

The reading has a precise operational meaning, and it is the meaning the title intends. The fine-tuning measure is the magnitude of the log-gradient of the observable in parameter space (Eq.~\ref{eq:sr}), and the island half-width $\epsilon \approx \ln 2 / \Delta$ converts it into the fractional range of the most sensitive parameter compatible with the observed abundance \citep{Profumo2026PBH}. A large $\Delta$, a small $\epsilon$, means that fixing the relic abundance to its measured value resolves the parameter sharply; a small $\Delta$, a large $\epsilon$, means the same measurement leaves the parameter loosely determined. The fine-tuning number is, read literally, a statement about measurement resolution: how precisely does the constraint $\Omega = \Omega_{\rm DM}$ determine the underlying parameter?

This is the legitimate content of naturalness, and it is an informational quantity in the strict sense. A stiff direction in the distinguishability geometry is one the data resolve; a sloppy direction is one they leave open \citep{Profumo2026ISR}. A ``fine-tuned'' parameter is one that the abundance observation, by fixing $\Omega$, resolves to high fractional precision; a ``natural'' one is a parameter the same observation barely constrains. The error of the tradition is to read this resolution statement as a plausibility statement: to move from ``the observed abundance pins this parameter to one part in $10^4$'' to ``a universe with this parameter is improbable or contrived.'' The first is a fact about the inferential leverage of a measurement on a map; the second is a claim about the world the measurement does not support. The ``borrowed authority''  licenses the move from the first to the second, and the informational reading blocks it.

\subsection{The embedded observer and the inaccessible ensemble}
\label{sec:embedded}

Why is the representational shell representational, and why does the invariant core survive? The companion account answers in terms of the situation of the observer. Cosmological objectivity is the objectivity available to observers embedded within the universe they study, constrained by causal horizons, by the finitude of accessible modes, and by the fact that only one realization of the relevant processes can be observed \citep{Profumo2026ISR, Bousso2002}. What counts as invariant is what remains stable across admissible representations for such observers.

This is the deepest reason the prior tuning of Section~\ref{sec:axion} is not an invariant. To convert a local sensitivity into a probability one needs a distribution over an ensemble: a prior over the misalignment angle, a measure over the parameter values some collection of universes would realize. For cosmology that ensemble is not accessible. We observe one universe, one value of each parameter, and no measurement pins down the distribution from which it was drawn, if such a distribution is even well posed. An embedded observer may of course adopt a prior; the point is that no measurement available to such an observer makes one prior uniquely objective, whereas the same observational resources fix the local sensitivity outright, recoverable from the abundance map alone. The invariant core of naturalness is the part an embedded observer can ground; the representational shell is the part no observational resource within the universe can render uniquely objective, since doing so would require a view of the ensemble of universes no such observer has. The invariant-versus-representational split of Section~\ref{sec:invariant} is, at bottom, the accessible-versus-inaccessible split of the embedded-observer epistemology.

Two objections sharpen rather than soften the claim. The first is that physics reasons about unsampled ensembles routinely; statistical mechanics assigns probabilities over microstates no one observes one by one. But the statistical-mechanical ensemble is not ungrounded: its measure is fixed by the dynamics, the Liouville measure preserved under Hamiltonian flow, and borne out by the repeated, exchangeable subsystems a macroscopic body supplies. A prior over a fundamental cosmological parameter has neither feature; no dynamical principle singles out the measure, and there is no ensemble of exchangeable universes to sample. The analogy breaks at exactly the point that would be needed to make the prior objective. The second objection is that cosmology has programs designed to supply the missing measure: eternal inflation and the string landscape set out to furnish a distribution over the values a parameter takes across regions or vacua \citep{BoussoPolchinski2000, Susskind2003}. They do, and their record is the clearest evidence for the present claim rather than against it. The predictions of such a measure depend on how the infinities of eternal inflation are regulated; different regulators yield different and sometimes observationally excluded answers, and no agreed principle selects among them, so that the program's own reviews record that the theory stays incomplete until the measure is fixed \citep{Freivogel2011}. This is not a measure that grounds the prior; it is the ungrounded prior reappearing one level up, as a choice of regulator standing in for a choice of distribution. The attempt to supply the ensemble relocates the gap and displays it; the inaccessible quantity remains inaccessible.

\subsection{The realism the account claims: structural and epistemic, not instrumental}
\label{sec:realism-status}

Two questions about the realism of this position must be met directly. The first asks whether the account is realist at all, given that the metric of Eq.~\eqref{eq:fisher} is scaled by an experimental error bar; the second asks, granting that it is, whether the realism is epistemic or ontic.

Take the error bar first. The Fisher metric pulled back from the Planck likelihood carries an overall factor $1/\sigma_\Omega^2$, so the metric tensor depends on the fractional precision $r$ of the measurement that induces it; as the abundance is measured more sharply, $r \to 0$, and the metric changes. If the geometry that carries the objective content changes with the state of our instruments, the worry runs, that content is an artifact of technology. The reply is that $r$ sets the unit of the geometry, not its shape. Rescaling $\sigma_\Omega$ multiplies every length by a common factor; it does not alter which directions are stiff and which sloppy, does not change the ratios of the eigenvalues, and does not move a construction from one universality class to another, because the class is fixed by the analytic form of the map and the band boundaries by physical scale ratios, neither of which contains $r$. What $r$ fixes is the representational shell in its quantitative form, the absolute resolution and the number $\Delta_{\rm SR}/r$; what improves as $r \to 0$ is the sharpness with which the fixed class boundaries are resolved, not the boundaries themselves. A position on which successive measurements resolve an instrument-independent structure ever more finely, without overturning it, is structural realism about that structure, not instrumentalism about the instrument.

The second question must be answered without hedging, and it is also where the two labels in play must be reconciled. The companion account's ``informational structural realism'' could in principle be read ontically; the position defended here is its \emph{epistemic} restriction. The defense against Newman in Section~\ref{sec:positive-thesis} establishes that an embedded observer's access to a cosmological parameter has a definite, measured, non-arbitrary structure; it secures a claim about epistemic access, not the claim that the parameter is nothing over and above that structure. The position is therefore epistemic structural realism about naturalness: the universality class is the structure of what an embedded observer can know of the parameter through the abundance, the limit within which the abundance measurement can resolve it at all, and the objective content of a naturalness verdict is exhausted by that structure. Whether the parameter is, in addition, ontologically nothing but structure is a question the account neither asserts nor requires.

It declines the ontic upgrade for a reason already in hand. To certify that a parameter is exhausted by what every admissible representation preserves, one would need a standpoint outside those representations from which to verify that nothing is left over; that standpoint is the view of the ensemble of universes that Section~\ref{sec:embedded} argued an embedded observer cannot occupy. The same inaccessibility that makes the prior representational makes the ontic thesis unreachable. The upgrade is also unnecessary: any feature of the parameter that no admissible representation preserves is, by construction, one no measurement available to the observer can reach, and so one that plays no part in physical inference. For inquiry conducted from within the universe the invariant is the whole of the objective content there is; the metaphysical question whether reality is ultimately structural can be left open without leaving anything in physics undetermined. This is the sense in which the position is informational structural realism in its epistemic form, rather than either a metaphysics of pure structure or an instrumentalism about appearances \citep{Profumo2026ISR}: it locates objective content in cross-representational informational invariants, and stops there, because the standpoint from which one could say more is exactly the one an embedded observer is denied.

\subsection{What naturalness is for}
\label{sec:what-for}

The account is constructive, not only corrective. If naturalness is the distinguishability geometry of the abundance map, its proper use is not to rank candidates by a plausibility it cannot supply, but to classify how sharply the cosmological observable constrains each candidate's parameters. A Class III map is one in which the measured abundance resolves the parameters tightly; a Class I map is one in which it scarcely resolves them at all. This is genuine and useful information: it tells us, for each candidate, how much inferential leverage the relic-abundance constraint carries, and therefore how much a future improvement in that constraint would teach us about the underlying physics. The reframing is most needed in the sector this paper has chosen. In the electroweak hierarchy the autonomy account supplies an independent grounding, and the informational reading is one defensible account among others; in the gravitational and cosmological sector, where Section~\ref{sec:gravity} argued that neither the symmetry nor the decoupling notion has purchase, the informational reading is not one option among several but what is left once the borrowed authority is withdrawn. There, naturalness is either an informational invariant of the distinguishability geometry, or nothing more than the aesthetic residue of a tradition that has outlived its grounds.

\section{Locating the rivals: autonomy of scales, post-naturalness, and non-empirical confirmation}
\label{sec:williams}

The positive account must answer to three rival positions: the autonomy-of-scales account it extends, the candid reassessment of naturalness from within the model-building tradition, and the program of non-empirical theory confirmation that offers naturalness its strongest rescue. The aim is not to defeat them but to locate them: to show where each is right, and where what it leaves unaddressed is what the informational reading supplies.

\subsection{Autonomy of scales: the grounding the informational reading extends}
\label{sec:williams-extend}

Section~\ref{sec:three-concepts} granted the autonomy account its home ground, and Section~\ref{sec:gravity} marked where that ground gives out. The relationship can now be stated exactly. Williams's notion and the informational notion are not competitors; the second generalizes the measurement the first performs, while conceding to the first a normative expectation the second does not supply.

Autonomy of scales is a statement about the map from high-energy inputs to low-energy observables: a theory is autonomous when its low-energy observables are insensitive to the ultraviolet, and the hierarchy problem is the failure of that insensitivity for the Higgs mass \citep{Williams2015}. In the language of Section~\ref{sec:isr-positive} this is a statement about the distinguishability geometry of a particular map: autonomy is sloppiness in the ultraviolet directions, and a violation of autonomy is stiffness in an ultraviolet direction, the sharp sensitivity of a low-energy observable to a high-energy input. The autonomy notion is therefore a special case of the informational notion, the case in which the parameters are ultraviolet inputs and the observable is a low-energy quantity. The informational reading is the genus; autonomy of scales is the effective-field-theory species.

The autonomy account nonetheless supplies something the informational reading, taken alone, does not. It supplies a reason to \emph{expect} sloppiness in the ultraviolet directions: decoupling is a generic feature of effective field theories, so a low-energy observable's insensitivity to the unknown ultraviolet is not merely measured but anticipated \citep{AppelquistCarazzone1975, Williams2015}. The link between this expectation and the autonomy of low-energy domains, and whether it suffices to justify a naturalness requirement, have been developed and contested elsewhere \citep{Wallace2019, Bain2019}. The informational reading classifies the geometry; the autonomy account adds a grounded expectation about what the geometry should be. On the electroweak home ground I grant the autonomy account this surplus without reservation.

The relocation of Section~\ref{sec:gravity} is now precise. In the gravitational and cosmological sector the autonomy account's measurement still translates, but its expectation does not. There is no decoupling theorem across the Planck scale to ground an expectation about the cosmological constant; there is no analog of generic decoupling for the abundance map of a dark matter candidate, whose structure is fixed by cosmological dynamics rather than by integrating out heavy modes. Williams's account ports to the gravitational sector only where a candidate inherits the electroweak hierarchy, as a supersymmetric relic inherits the Higgs sensitivity through a shared breaking sector; for the cosmological constant, for the thermal-relic coincidence, and for candidates with no tie to the hierarchy, the expectation is unavailable. What remains is the measurement: the distinguishability geometry of the map. The informational reading is thus not a rival to autonomy of scales but its continuation into the sector where the autonomy grounding runs out and only the geometry is left.

\subsection{Post-naturalness as evidence of over-trust}
\label{sec:giudice}

The autonomy account has a further consequence its own proponents have drawn. If naturalness in the electroweak sector is the autonomy expectation, then its empirical disappointment, the absence of new physics at the scale the expectation pointed to, is a defeasible structural expectation meeting a contrary result. That is an ordinary scientific outcome: the expectation was reasonable, the data did not bear it out, the expectation is revised. Williams himself frames the Large Hadron Collider null result in just these terms \citep{Williams2015}.

The reaction within the model-building tradition has not had the character of ordinary revision. The reassessment that followed the null results speaks of a crisis, of a post-naturalness era, of disorientation about what beyond-the-Standard-Model theorizing is now for \citep{Giudice2013, Giudice2017}. I take that reaction, offered candidly from inside the program, as evidence for the thesis of this paper. The disorientation is disproportionate to the defeat of a defeasible structural expectation; a reasonable expectation that fails to be realized does not, by itself, unsettle the purpose of a field. The disproportion is the tell. Naturalness was carrying more than the autonomy content could account for: it was carrying the affective certainty of the older tradition, the conviction that {\em the world owes us the absence of unexplained coincidence}, and it is the defeat of that surplus conviction, not of the structural expectation, that registered as a foundational disappointment. Read this way, the post-naturalness literature is not a counterexample to the borrowed-authority thesis but its most candid confirmation, supplied by practitioners describing the gap between what naturalness could ground and what its failure cost.

\subsection{Non-empirical confirmation and the limits of the rescue}
\label{sec:dawid}

The strongest defense of naturalness does not depend on autonomy of scales at all. It grants that naturalness operates without direct empirical test and argues that this is no defect, because there are rational, non-empirical grounds for theory assessment. On Dawid's account, the trust theorists place in an untested theory can be reconstructed as a form of confirmation: through the observation that no alternative has been found, through a meta-induction over the past success of theories with the relevant virtues, and through unexpected explanatory coherence \citep{Dawid2013}. If naturalness is such a non-empirical criterion, then its authority is earned in the way these arguments describe, not borrowed, and the central claim of this paper is wrong.

This is the objection to take most seriously, and the answer has two parts. The first concerns what Dawid's arguments do. They reconstruct the rationality of a trust already in place; they take the existence and direction of that trust as input and ask whether it can be rationally underwritten. They are, by design, silent on a different question: why the trust took this particular form. The trust naturalness commands is quantitatively rigid, fixed to a threshold; it is historically specific, arising in a definite period and tied to definite model-building cultures; and it is so resistant to revision that, as Section~\ref{sec:giudice} observed, its empirical failure produced a crisis rather than an update. The genealogy and the grip of that specific disposition are exactly what the present paper asks about, and exactly what a reconstruction of non-empirical rationality brackets. The two accounts answer different questions, and his does not displace mine.

The second part concerns the meta-inductive argument specifically, because it is the form a Dawidian defense would most naturally take: naturalness has a track record, and a meta-induction over it licenses continued trust. The track record is real, but it does not say what the argument needs. The canonical successes, among them most famously the bound on the charm-quark scale from the structure of the neutral-current sector, and the prediction of the scale of new physics from the electromagnetic pion mass splitting, are successes of technical naturalness in the sense of Section~\ref{sec:technical}: each is a symmetry-and-sensitivity argument grounded in a definite structural fact \citep{Giudice2008}. They are not successes of the quantitative fine-tuning measure read as a plausibility verdict, nor of the aesthetic disposition the genealogy traces. A meta-induction over the record therefore vindicates the notion that was already grounded and lends nothing to the notion the paper contests. The same disentangling disposes of two adjacent non-empirical defenses: the pragmatic case that naturalness has earned its standing as a heuristic for locating new physics \citep{Wells2015}, and the forward-looking case that a principle is justified by the research program it makes possible rather than by past confirmations \citep{Fischer2023}, are each strongest for the symmetry and autonomy notions that have in fact guided successful model-building, and weakest for the quantitative measure read as a plausibility verdict. Like Dawid's reconstruction, each reaches the grounded species of naturalness and not the borrowed one.

What Dawid's program can rationalize, then, is the invariant core of Section~\ref{sec:isr-positive}: the structural, non-aesthetic content of a naturalness judgment, the distinguishability geometry a non-empirical assessment could legitimately track. What it cannot rescue is the representational shell, whose grounding, as Section~\ref{sec:embedded} argued, would require a view of an ensemble of universes no embedded observer has. Non-empirical confirmation reaches the part of naturalness that was never the problem and stops exactly where the problem begins.

\section{Against the deflationary overcorrection}
\label{sec:deflation}

The genealogy and the positive account together invite a deflationary reading that must be distinguished from the present one, because it shares this paper's starting point and reaches the opposite destination. The deflationary critic agrees that naturalness has carried unearned authority and concludes that naturalness is therefore empty. The conclusion overshoots the premise: the critique is right about what it attacks and wrong about what it discards, and the difference is the invariant core of Sections~\ref{sec:invariant} and~\ref{sec:isr-positive}.

\subsection{What the critique gets right}
\label{sec:deflation-right}

The most rigorous form of the deflationary case is not the charge that naturalness is ugly aesthetics but a precise technical argument, and it must be credited in full. Arguments from naturalness in their probabilistic form treat the observed value of a parameter as improbable, and an improbability claim requires a probability distribution over the parameter; but no such distribution is given, and any attempt to supply one introduces an arbitrariness that conflicts with naturalness itself, since naturalness was meant to remove arbitrariness rather than relocate it into a prior \citep{Hossenfelder2018}. The same demand, that a fine-tuning argument is only as well posed as the measure it presupposes, has been pressed independently as a criterion separating cogent fine-tuning arguments from ill-defined ones \citep{Grinbaum2012}. Without a specified distribution the improbability reading is ill-defined; with one it is self-undermining.

This paper does not contest that argument; it reaches the same verdict by another route. Section~\ref{sec:embedded} argued that the prior over a cosmological parameter is exactly the quantity an embedded observer cannot ground. The probabilistic reading of naturalness, the reading that treats a tuning figure as an improbability, is the representational shell of Section~\ref{sec:invariant}, and on its ill-definition the deflationary critic and the present account agree. The disagreement is not about the shell.

\subsection{Where the critique overreaches}
\label{sec:deflation-overreach}

The deflationary case moves from the ill-definition of the probabilistic reading to a verdict on naturalness as such: that it amounts to aesthetic preference with little record of usefulness, and that efforts to resolve naturalness problems are ``a waste of time'' \citep{Hossenfelder2018}. This inference is valid only if the probabilistic reading exhausts naturalness. It does not, and the measure argument that defeats the probabilistic reading has no purchase on what remains.

What remains is prior-independent. The local sensitivity of Section~\ref{sec:measures} is a derivative of the abundance map, defined without any distribution over parameters; the universality class of Section~\ref{sec:invariant} is an equivalence class under reparametrization, again involving no distribution. Neither is an improbability claim, and the argument that improbability claims require an ungrounded prior therefore touches neither. The overreach is enabled by the conflation of Section~\ref{sec:three-concepts}: by treating naturalness as a single object, the critic lets an argument that defeats only the probabilistic reading stand as a refutation of naturalness entire. Disentangled, the argument has no purchase on technical naturalness, which is a statement about symmetry and invokes no probability; none on the autonomy of scales, which is a statement about decoupling; and none on the informational invariant, which is a statement about map structure.

The verdict that naturalness has little track record of usefulness is where the conflation does the most work, because it holds of the probabilistic reading and fails of the notion that does have a record. The historical successes noted above are successes of technical naturalness; they are not improbability arguments, and they are exactly the cases the measure critique leaves standing. A verdict of uselessness drawn from the failure of the probabilistic reading cannot be transferred to the symmetry reading that supplied the successes. The residue the deflationary critique discards is not an aesthetic remainder but the structural content the rest of this paper has isolated: the grounded notions of Section~\ref{sec:three-concepts}, and the informational invariant of Section~\ref{sec:isr-positive} that gives a naturalness judgment its legitimate, prior-free, inferential-resolution meaning.

\subsection{The middle position}
\label{sec:deflation-middle}

The deflationary critic and the naive realist about naturalness make opposite errors about the same number. The realist reifies it, reading a representation-dependent coordinate as a deep fact about the world; the critic, having shown the coordinate cannot bear that reading, denies that any objective content remains. The middle position is the one this paper defends: the number is representational, and the class is real. Both halves are needed, and each rival holds one while dropping the other.

A last observation is offered as explanation, not argument, because the case against the realist reading has already been made on structural grounds and the genealogy adds nothing to it. The deflationary critique is candid that fine-tuning still strikes its author as ugly and natural theories as more beautiful, even as the argument denies that the feeling tracks anything in the world \citep{Hossenfelder2018}. That admission is not a lapse and not evidence; the work of showing that the feeling has no objective correlate was done structurally in Sections~\ref{sec:gravity}--\ref{sec:isr-positive}, and would stand whether or not anyone still felt the pull. What the admission supplies is the thing the genealogy was introduced to explain. A disposition whose grounds have been withdrawn {\em continues to be felt, with undiminished force, by the very people who have withdrawn them}; the persistence of the conviction in its most rigorous critic is the borrowed authority of Section~\ref{sec:genealogy} exhibited where one would least expect it. The genealogy does not refute the realist, who is refuted elsewhere; {\em it explains why the refuted view goes on feeling true}.

\section{Conclusion: what naturalness measures}
\label{sec:conclusion}

This paper began from a discrepancy: naturalness is invoked as an objective constraint on physical theories, yet its verdicts move when conventions move. The resolution has been to separate the invariant core of a naturalness judgment from its representational shell. The core is the universality class of the map from parameters to observables, invariant under admissible changes of parametrization and measure convention and independent of any prior over parameter space; the shell is the tuning number and the conventions that fix it. The number is a coordinate on the invariant content, not the content; and the felt authority of a naturalness verdict, the conviction that a large tuning figure reveals a contrived world, attaches to the shell.

That authority is inherited. The disposition to read simplicity and beauty as guides to truth has persisted for centuries while the grounds offered for it have been rewritten by each epoch in turn, and modern naturalness has inherited the authority of that tradition without inheriting any one of its successive justifications. The gap between the authority and the grounds is widest in the gravitational and cosmological sector, where the effective-field-theory account that grounds naturalness for the Higgs has no purchase: the cosmological constant, where no decoupling statement crosses the Planck scale; dark energy, where the secure content is a distance--redshift relation and the tuning attaches to one revisable encoding of it; and the relic abundance of dark matter, where a uniform analysis shows that fine-tuning tracks the analytic structure of the abundance map and not the nature of the candidate, that the tier is invariant while the number is not, and that gravitational relics span every tier \citep{Profumo2026PBH}. It is there, where the borrowed authority does the most work and the structural grounding the least, that the distinction between core and shell is sharpest.

The title's question can now be answered. {\em Naturalness measures the distinguishability geometry of the representations through which physics encodes observation}. Read literally, a fine-tuning number states how sharply a measurement, by fixing an observable, resolves an underlying parameter; it is a statement about inferential resolution, not about the contrivance of the world. This is genuine and useful content: it says how much leverage a constraint carries, and how much a sharper measurement would teach. What it does not say, and has been taken to say, is that a finely resolved parameter belongs to an improbable or inelegant universe. That further claim requires a view of an ensemble of universes that an observer embedded in one of them cannot have, and it is the part of naturalness that dissolves under examination, leaving the structural part intact.

The three rivals, located rather than refuted, each mark an edge of this account: the autonomy-of-scales account supplies, on its electroweak home ground, a grounded expectation the informational reading generalizes; the post-naturalness reassessment testifies, from inside the program, to the gap between what naturalness could ground and what its failure cost; and the deflationary critique correctly condemns the probabilistic shell while wrongly discarding the structural core along with it. Progress, on this account, is not what the tradition supposed. It is not convergence toward an ever more beautiful ontology, but the refinement of informational structure: as the relic abundance and its correlated observables are measured more precisely, the equivalence classes are resolved more finely, and what improves is the resolution of the distinguishability geometry accessible to observers situated within the universe they study.

Naturalness does not measure how contrived the world is; it measures how sharply what an embedded observer can see constrains the representations available for seeing it. The elegance we took for a feature of the world is, on this account, a feature of our maps, and of the geometry by which the world holds them to account.

\section*{Acknowledgments}
This work was supported in part by the U.S.\ Department of Energy,
Office of Science, Office of High Energy Physics, under Award
Number DE-SC0010107.
\appendix
\section{Derivation of the single-exponential identity}
\label{app:single-exp}

This appendix derives Eq.~\eqref{eq:single-exp-identity} and exhibits its independence from the microphysics that supplies the exponential. The result is established in the companion analysis \citep{Profumo2026PBH}; it is reproduced here so that the philosophical claim it supports does not rest on an unstated computation. The derivation has two steps. The first fixes the value of the exponent on the observed-abundance contour by cosmological bookkeeping alone; the second relates that value to the sensitivity measure.

\paragraph{The single-exponential class.}
A construction belongs to the single-exponential class when its present dark matter density can be written
\begin{equation}
\Omega(x) \;=\; \Omega_{\rm natural}\, e^{-f(x)},
\label{eq:app-class}
\end{equation}
with $\Omega_{\rm natural}$ the order-unity density the construction would yield with the suppression switched off, $f(x) \gg 1$ a model-dependent exponent, and $x$ the parameter that controls it. The defining feature of the class is that a single exponential factor dominates the departure of $\Omega$ from its natural value; the function $f$ may be a nucleation action, a Boltzmann exponent, or a collapse suppression, and the derivation does not depend on which.

\paragraph{Step one: the exponent on the contour is a clock.}
Let the relic be a matter-like component formed at temperature $T_{\rm form}$, carrying at formation a fraction $\beta \equiv \rho_X / \rho_{\rm rad}$ of the radiation energy density. Because the component redshifts as $\rho_X \propto a^{-3}$ against $\rho_{\rm rad} \propto a^{-4}$, the ratio grows as $\rho_X / \rho_{\rm rad} \propto a \propto 1/T$, the last step using entropy conservation $aT \simeq \text{const}$ up to the slowly varying $g_{*s}$. Matter--radiation equality is the epoch at which $\rho_{\rm DM}/\rho_{\rm rad} = 1$, by definition, so tracing the observed dark matter back to formation gives
\begin{equation}
\beta_{\rm obs} \;=\; \frac{\rho_X}{\rho_{\rm rad}}\bigg|_{\rm form}
\;=\; \frac{a_{\rm form}}{a_{\rm eq}}
\;=\; \frac{T_{\rm eq}}{T_{\rm form}} \;\ll\; 1 .
\label{eq:app-beta}
\end{equation}
A natural formation, one that converts an order-unity fraction of the bath, has $\beta_{\rm natural} = O(1)$; the observed abundance requires the suppression $\beta_{\rm obs}/\beta_{\rm natural} = T_{\rm eq}/T_{\rm form}$. Since the present density is proportional to the formation fraction times fixed redshift factors, $\Omega_{\rm natural}/\Omega_{\rm DM} = \beta_{\rm natural}/\beta_{\rm obs} = T_{\rm form}/T_{\rm eq}$. Imposing $\Omega = \Omega_{\rm DM}$ in Eq.~\eqref{eq:app-class} therefore fixes the exponent to
\begin{equation}
f\big|_{\Omega = \Omega_{\rm DM}}
\;=\; \ln\!\left( \frac{\Omega_{\rm natural}}{\Omega_{\rm DM}} \right)
\;=\; \ln\!\left( \frac{T_{\rm form}}{T_{\rm eq}} \right).
\label{eq:app-clock}
\end{equation}
The value of the exponent on the contour is set by the ratio of two physical scales, the formation scale and the scale of equality, and by nothing in $f$ itself. This is the sense in which it is a cosmological clock: the microphysics chooses where in parameter space the contour lies, but the height of the exponent on that contour is fixed by redshift bookkeeping.

\paragraph{Step two: the sensitivity equals the exponent.}
The Barbieri--Giudice measure is $\Delta = |\partial \ln \Omega / \partial \ln x| = |\partial f / \partial \ln x|$, using Eq.~\eqref{eq:app-class} and the constancy of $\Omega_{\rm natural}$. In an admissible parametrization (Section~\ref{sec:tier-invariant}), where $x$ is the input scale that sets the exponent, $f$ is to leading order homogeneous of degree one in $\ln x$ near the contour, so $|\partial f / \partial \ln x| = c\, f$ with $c = O(1)$. The condition amounts to $f$ behaving as a power $f \propto x^{c}$ near the contour; the canonical example is a mass parameter $x = M$ entering a Boltzmann suppression $f \simeq M/T_\star$ at a formation temperature $T_\star$ set independently of $M$, which gives $\partial f / \partial \ln M = M/T_\star = f$ and hence $c = 1$. Combining with Eq.~\eqref{eq:app-clock},
\begin{equation}
\Delta\big|_{\Omega = \Omega_{\rm DM}}
\;=\; c \, \ln\!\left( \frac{T_{\rm form}}{T_{\rm eq}} \right),
\qquad c = O(1),
\end{equation}
which is Eq.~\eqref{eq:single-exp-identity}. With $T_{\rm eq} \simeq 0.8~{\rm eV}$ and formation scales ranging from $\sim{\rm MeV}$ to the grand-unified scale, the logarithm runs from about $14$ to about $50$, the Class II band. The $O(1)$ coefficient $c$ is where the residual model dependence lives; the order of magnitude is not model-dependent, because it is the logarithm in Eq.~\eqref{eq:app-clock}.

\paragraph{Robustness and the migration to Class III.}
The derivation used only Eq.~\eqref{eq:app-class} and the redshift scaling of a matter-like relic, so it holds across constructions that share the single-exponential form regardless of the mechanism: a first-order phase transition (with $f$ the nucleation action, independent of the detailed $S(\alpha)$), an early matter-dominated era, and coannihilation (with $T_{\rm form}$ the freeze-out temperature) all satisfy it. The one place the result can fail is the place the main text flags. If the true exponent grows faster than log-linearly in the controlling parameter, as a super-exponential collapse probability does, then $|\partial f / \partial \ln x| \gg f$ and $\Delta$ exceeds the bound of Eq.~\eqref{eq:single-exp-identity}, carrying the construction into Class III. This is consistent with the account rather than an exception to it: the invariant content is the analytic form of the map, so describing that form more faithfully can move the class, while the identity remains exact within the single-exponential class it characterizes.

\bibliographystyle{plainnat}

\bibliography{references}

\end{document}